# FeOOH instability at the lower mantle conditions

E. Koemets[1], T. Fedotenko[2], S. Khandarkhaeva[1], M. Bykov[1,3], E. Bykova[3], M.Thielmann[1], S. Chariton[1], G. Aprilis[2], I. Koemets[1], H.-P. Liermann[3], M. Hanfland[4], E.Ohtani[5], N. Dubrovinskaia[2], C. McCammon[1], L. Dubrovinsky[1]

1. Bayerisches Geoinstitut, University of Bayreuth, D-95440 Bayreuth, Germany
2. Material Physics and Technology at Extreme Conditions, Laboratory of Crystallography, Universität Bayreuth, D-95440 Bayreuth, Germany
3. Photon Science, Deutsches Elektronen-Synchrotron, D-22607 Hamburg, Germany
4. ESRF-The European Synchrotron CS40220 38043 Grenoble Cedex 9 France
5. Department of Earth Science, Graduate School of Science, Tohoku University, Sendai 980-8578, Japan;

**Goethite, α-FeOOH, is a major component among oxidized iron species, called rust, which formed as a product of metabolism of anoxygenic prokaryotes** (*1, 2*) **inhabiting the Earth from about 3.8 billion years (Gy) ago until the Great Oxidation Event (GOE) of about 2.5 Gy ago. The rust was buried on the ocean floor** (*1, 2*) **and had to submerge into the Earth mantle with subducting slabs due to the plate tectonics started about 2.8 Gy ago** (*3*)**. The fate and the geological role of the rust at the lower mantle high-pressure and high-temperature(HPHT) conditions is unknown. We studied the behavior of goethite up to 82(1) GPa and 2300(100) K using** *in situ* **synchrotron single-crystal X-ray diffraction. At these conditions, corresponding to the coldest slabs at the depth of about 1000 km, α-FeOOH decomposes to various iron oxides ($Fe_2O_3$, $Fe_5O_7$, $Fe_7O_{10}$, $Fe_{6.32}O_9$) and an oxygen-rich fluid. Our results suggest that recycling of the rust in the Earth mantle could contribute to oxygen release to the atmosphere and explain the sporadic increase of the oxygen level before the GOE linked to the formation of Large Igneous Provinces**(*4*)**.**

Water or water-bearing species have a strong impact not only on life on our planet but also on numerous processes in Earth's interiors (*5, 6*). Presence of water affects various properties of mantle minerals and causes different global phenomena such as arc volcanism and plate tectonics (*7–9*). Mechanisms of water circulation between geospheres are crucial for understanding our planet's geodynamics and geochemistry but still remain debated and discussed actively (*6*). Recent studies suggest that iron-bearing minerals present in Banded Iron Formations (BIFs), such as goethite, could transport some quantities of water to the deep Earth interiors with subducting slabs (*10–13*). Hu et al. (*12, 13*) and Nishi et al. (*11*) reported that goethite remains stable in the sinking slab until it reaches the bottom part of the lower mantle. By means of *in situ* powder XRD it was revealed that at pressures corresponding to the depths of ~1500-1800 km and at moderately high temperatures goethite undergoes a phase transition to form a stable pyrite-type phase $FeO_2H_x$ with $0 \leq x \leq 1$ (Py-phase) (*11–13*). Therefore goethite and its HP polymorph are considered to be candidates for water and/or hydrogen transfer to the lower mantle and the core-mantle boundary (*10*). Here we present a detailed study of the behavior of FeOOH at conditions of the subducting slab using advanced *in situ* single-crystal X-ray diffraction technique in laser-heated diamond anvil cells (LH-DACs).

The behavior and stability of hydrous phases at certain depths depend on the pressure – temperature conditions experienced by the materials. While the pressure profile of the Earth's interiors is well defined (*14*), the temperature distribution within the subducting slabs remains the subject of discussions (*15*). Thus, a reliable estimation of slab's geotherm is the key question in the study of the stability of goethite and its transportation to the lower mantle. Recent studies on FeOOH stability (*11*) were performed at a "cold slab" conditions suggesting that temperature inside the slab barely reaches 1700 K at 120 GPa (ca. 2700 km depth). This estimation of the slab geotherm was based on the extrapolation of data of Eberle et al. (*16*) who modeled slabs' behavior to the depth of only about 800 km.

Taking into account that a simple extrapolation of a slab's temperature from the depth of 800 km to the Core-Mantle boundary depths of ~2900 km is problematic, we estimated the low boundary of a temperature distribution in the slabs sinking into the Earth mantle by numerical modeling calculations (See *Supplementary Information* for details). According to the geodynamic analysis (*17*), average vertical velocity of the sinking slabs (as a component of angular slab immersion speed towards the Core, not to be confused with the plate speed) can hardly be higher than 1 cm/year ("cold slab"), but we also considered a case of 2 cm/year ("ultra-cold slab") to estimate the lowest possible temperatures across the slab. The results of the temperature estimation (Fig. 1) provide constraints on the PT conditions for LH-DAC experiments aimed at investigation of goethite stability in present work (see *Methods, Table S1*).

We performed a series of *in situ* synchrotron single-crystal XRD experiments in laser heated diamond anvil cells at pressures up to about 80 GPa and in a wide temperature range

(Fig. 1, Table S1 in *Supplementary Materials*). Heating of a goethite crystal to 1500(100) K at 40(2) GPa (run 1, Table S1), that corresponds to the PT conditions at ~1000 km depth of a cold slab, leads to a transformation of initial α-FeOOH to ε-FeOOH (space group *Pnnm*) and its partial decomposition resulting in formation of the $Rh_2O_3$-II type structured iron oxide ι-$Fe_2O_3$ (*18, 19*) (Fig. 1, Table S2, Fig. 2). At slightly higher temperature of 1600(100) K at 41(2) GPa (run 2, Table S1), the decomposition yields two iron oxides - ι-$Fe_2O_3$ and monoclinic $Fe_5O_7$ (*19*) (Fig. 1, Table S2, Fig. 2). Heating at 52(1) GPa and 1200K (run3.1, Table S1) resulted in the formation of three iron oxides - previously known $Fe_5O_7$ and η-$Fe_2O_3$ ($CaIrO_3$-type $Fe_2O_3$ (*19*)), and an additional hexagonal phase (Fig. 1, Table S2, Fig. 2). This phase, according to single crystal X-ray diffraction, has a chemical composition $Fe_{6.3}O_9$ and the hollandite ($NaLi_2Ru_6O_{12}$ (*20*))-like structure (Fig. 1, Fig. 2, Fig. S4 and supplementary information (SI)). Heating of another goethite crystal in the same DAC at 1550(100)K (run 3.2, TableS1) leads to the formation of η-$Fe_2O_3$ and $Fe_5O_7$, while no traces of $Fe_{6.3}O_9$ were detected (Fig. 1). Treatment of goethite (run 3.3, Table S1) at about 2000 K at 52(1) GPa leads to formation of a single iron oxide, orthorhombic HP-$Fe_3O_4$ (*19*) (Fig. 1, Table S2, Fig. 2). During experiments at 64(2) GPa we registered (run 4, Table S1) decomposition of FeOOH with a formation of η-$Fe_2O_3$ at 1500(100)K. On further iterative sample heating up to 2100(100)K (run 4, Fig. 1, Table S1) we observe formation of η-$Fe_2O_3$ and novel orthorhombic iron oxide, $Fe_7O_{10}$ (space group *Cmcm*), (Fig. 2, Table S2 and SI). Upon laser heating at 81(2) GPa and 1500(100) K (run 5, Fig. 1) goethite transforms into cubic $FeO_2H_x$ as it was reported earlier (*21*) and partially decomposes into η-$Fe_2O_3$ and $Fe_{6.32}O_9$. Cubic $FeO_2H_x$ has the unit cell volume, which is larger than that of hydrogen-free $FeO_2$ (86.94(8) Å$^3$ vs 81.7(2) Å$^3$ (*13*) correspondingly), so that the estimated amount of hydrogen incorporated into the structure is x=0.55(5) if the one considers the $FeO_2$ and ε-FeOOH as the end-members (*22*).

Summarizing our experimental observations, we have found that goethite is not stable even at relatively low temperatures (see Fig. 1, Table S1). In the pressure range of 40-80 GPa and at temperatures of 1200-2100 K, it decomposes forming various iron oxides: ι-$Fe_2O_3$, η-$Fe_2O_3$, HP-$Fe_3O_4$, and $Fe_5O_7$, previously reported in the literature (*19*) and $Fe_7O_{10}$ and $Fe_{6.32}O_9$ found in the present work. Due to the experimental conditions (LH-DACs), the question of thermodynamic stability and equilibrium of the phases observed at certain P-T conditions remains unclear. Nevertheless, our observations unambiguously suggest that the oxidation-reduction reaction occurs producing $Fe^{2+}$

In order to consider a possibility of the presence of hydrogen in the studied mixed-valence iron oxides, we analyzed their densities based on the XRD experiments data (Fig. 3). Incorporation of hydrogen into the structure drastically affects the density, for example, at 40(2) GPa the density of hydrogen-bearing $\varepsilon$-FeOOH (5.34 g/cm$^3$, (*23*)) is significantly lower than that of hydrogen-free $FeO_2$ (6.59 g/cm$^3$, (*13*)). The density of $\iota$-$Fe_2O_3$ (6.25 g/cm$^3$) synthesized from goethite at 40(2) and 41(2) GPa matches the value for hydrogen-free iron oxide (*19*) at the same pressure. Synthesized at 41(2) GPa, $Fe_5O_7$ can be considered as a mechanical mixture of FeO (*24*) and $Fe_3O_4$ (*25*) so its density should be an average of FeO and $Fe_3O_4$. If it is hydrogen-free, its density should be equal to 6.50 g/cm$^3$ that matches well the experimental value of 6.48 g/cm$^3$ for $Fe_5O_7$. The density analysis allowed us to conclude that iron oxides obtained in the present work do not contain any significant amount of hydrogen (Fig. 3). Densities of the previously unknown phases, $Fe_7O_{10}$ and $Fe_{6.32}O_9$, match the linear composition-density trend established for various hydrogen-free iron oxides at appropriate pressures (Fig. 3). Notably, synthesized at relatively high temperatures $Fe_7O_{10}$ (Fig. 3, Table S2) is even slightly denser than one could expect from the trend (Fig. 3). The lesser density of $Fe_{6.32}O_9$ might indicate that the material could contain some amount of hydrogen (located, for example, in channels of the structure, Fig. 2), but even if present, the amount of hydrogen in the $Fe_{6.32}O_9$ should be very small. As this phase was not observed at temperatures higher than 1500 K (Fig. 1), it is unlikely to carry water into the deep Earth interiors.

Our observations may have serious implications for understanding of water (H-bearing component(s) of the complex oxides under consideration) transport in the lower mantle and geochemistry of the Earth interior. As we found that FeOOH is not stable even at conditions of the "super-cold" slab at the depth of about 1000 km, iron oxyhydroxide is unlikely to be the carrier of water to the deep Earth regions, such as the deep lower mantle and the core-mantle boundary, under realistic slab subduction conditions.

Observed appearance of iron(II) bearing oxides (particularly, $Fe_5O_7$, $Fe_7O_{10}$, $Fe_{6.32}O_9$, etc.) evidences the release of free oxygen by analogy with $Fe_2O_3$ HP-HT treatment reported by Bykova et al. (19). Considering that $Fe_3O_4$ has the highest amount of $Fe^{2+}$ among the observed compounds, schematically this process can be described as:

12FeOOH -> 4Fe$_3$O$_4$ + 6H$_2$O + O$_2$.   [2]

Since this process for FeOOH is observed at significantly milder pressures and temperatures than required for decomposition of pure $Fe_2O_3$ (19), water either directly involved or facilitate reduction of ferric iron. Thus, we can suggest that water- and ferric-bearing phases become unstable upon heating at high pressure due to self-redox reactions.

Our experiments demonstrate that the depth of about 1000 km poses a "chemical border", below which water- and ferric-iron-bearing compounds produce an oxidizing fluid. Obviously, different phases of FeOOH are candidates to be a possible source of such a fluid, but the amount of iron(III) oxyhydroxides is currently negligibly small on the surface of our planet to consider them as a global geochemical component. However, situation might have been different in the past: before the advent of oxygenic photosynthesis, the Earth's biosphere was driven by anaerobic metabolisms, starting from about 3.8 Gyr ago, as evidenced by the earliest preserved marine sediments (1). During this period, iron(III) oxyhydroxides were probably a common product of vital functions of iron-oxidizing phototrophs (1). Schematically, this process in early oceans could be illustrated (1) by the following reaction:

4Fe$^{2+}$ + CO$_2$ + 7H$_2$O -> 4FeOOH + C·H$_2$O + 8H$^+$,   [1]

where "C·H$_2$O" denotes organic materials produced in biochemical processes.

The amount of ferric iron produced and buried, as a result of ferrous iron oxidation due to anoxygenic photosynthesis, is estimated as 1.1·10$^{13}$ mol/year (1). If the predominant mineral form of ferric iron in sediments was FeOOH (called "rust" by some authors (26)), it would take less than 5 million years to accumulate the amount of oxygen present in the modern atmosphere.

When plate tectonics became active (between 3.2 and 2.5 Gyr ago, (*3*)), subduction brought sediments rich in iron oxyhydrates down to deep Earth's interiors. Even if some part of subducted FeOOH-bearing sediments would be consumed by metamorphic processes and reactions with surrounding mantle materials, the same arguments in favor of their preservation during subduction hold, as those valid for banded iron formations (BIFs) sinking down to core-mantle boundary(*19, 27*). Assuming the sinking speed was the same as in modern slabs (around 1 cm/yr., as a vertical component of slab subduction speed), the necessary time to access the depth of 1000 km was just 0.1 Gyr. Thus, significant part of FeOOH-enriched material produced by phototrophs might reach the "chemical border" and release oxygen-rich fluids. These oxidizing fluids travelling to the Earth surface might contribute to (or even be one of the causes of) the sequence of the events known as the Great Oxidation Event (GOE). As demonstrated by recent studies (*4*), GOE is a time-distributed process occurred about 2.5 Gyr ago and associated with a formation of large igneous provinces. Additionally, recent studies of the Xe geochemistry (*28–30*) suggest that massive mantle degassing occurred ~2.7-2.5 Gyr ago thus providing mechanism (and time frame) for the oxygen release from the lower mantle.

If our hypothesis is correct, formation of the oxygen-rich atmosphere on Earth was a result of a combination of processes in the early anaerobic biosphere, plate tectonics, and decomposition of the ferric iron oxyhydroxides in the Earth's mantle. Oxygen-based life as we know it today could be a result of the adaptation of early-living forms to rising oxygen content in the atmosphere due to the geological processes.

## Methods

### Sample preparation and DAC assembly

Single crystals of natural goethite with an average size of ~0.01 × 0.01 × 0.01 mm3 (*23*) were used in this study. In all runs, crystals were placed inside BX-90 diamond anvil cell (*31*) equipped with Boehler-Almax type anvils. We used pre-indented oil-free Re gaskets to ~20 μm thickness of ~20 μm. Sample chambers were drilled inside gaskets using the laser with a diameter < ½ diameter of an anvil's culets. The cells were loaded with Ne as a pressure-transmitting medium using the gas-loading system installed at the Bayerisches Geoinstitut. Pressure determination inside DAC was performed using diamond anvil Raman shift (*32*) and EOS of Ne (*33*). All the cells were cold compressed to the desired pressure and then goethite crystals were heated at different laser-heating systems. Further, the samples were investigated in-situ using synchrotron single-crystal XRD. For more technical details, please see supplementary table ST1.

### X-ray diffraction

The single-crystal X-ray diffraction experiments were conducted on the ID15B beamline at the European synchrotron radiation facility (ESRF), Grenoble, France (MAR555 detector, λ=0.4126–0.4130 Å); on the extreme conditions beamline P02.2 at PETRA III, Hamburg, Germany (PerkinElmer XRD1621 flat panel detector, λ=0.2898–0.2902 Å). The X-ray spot size depended on the beamline settings and varied from 4 to 30 μm, where typically a smaller beam was used for laser heating experiments. A portable double-sided laser heating system was used for experiments on ID15B (ESRF) to collect ex situ single-crystal X-ray diffraction. State-of-the art stationary double-side laser-heating set-up at P02.2 (PETRA III) has been used for *in situ* high temperature single-crystal X-ray diffraction along with temperature-quenched single-crystal X-ray diffraction. Laser spot size on a sample was covering the whole crystals and there were no measurable temperature gradients within the samples. In the case of prolonged heating experiments, the temperature variation during the heating did not exceed ±100 K. Pressures before and after laser-heating were calculated from the EOS of Ne (*33*). X-ray diffraction images were collected during continuous rotation of DACs typically from –38° to +38° on ω; while data collection experiments were performed by narrow 0.5° scanning of the same ω range. For the crystallographic information on the phases of present paper please see ST2.

### Data analysis

Whereas the starting material, a goethite single crystals, gives characteristic Bragg peaks in the diffraction pattern, after the laser-heating in solidified neon, we clearly observed well defined, sharp diffraction spots belonging to the multiple grains of other high-pressure phases, already known and novel. Using the Ewald$^{Pro}$ reciprocal space viewing tool for the CrysAlis$^{Pro}$ software (*34*), we were able to identify the diffraction spots belonging to certain domains, find their matrices of orientation and refine the unit cell parameters within subsequent intensities extraction (data integration). The crystal structures were solved using SHELXT software that employs a dual-space algorithm for the solution of a phase problem(*35*). General output of the structure solution program was a crystallographic position(s) of a heavy iron atom(s), while locations of oxygen atoms were assigned based on the analysis of residual electron density maps. Crystal structures were refined against single-crystal diffraction data using the JANA2006 software (see ST2 for crystallographic information on the phases of present paper). The general procedure of the analysis of a multigrain diffraction dataset is described in (*36*).

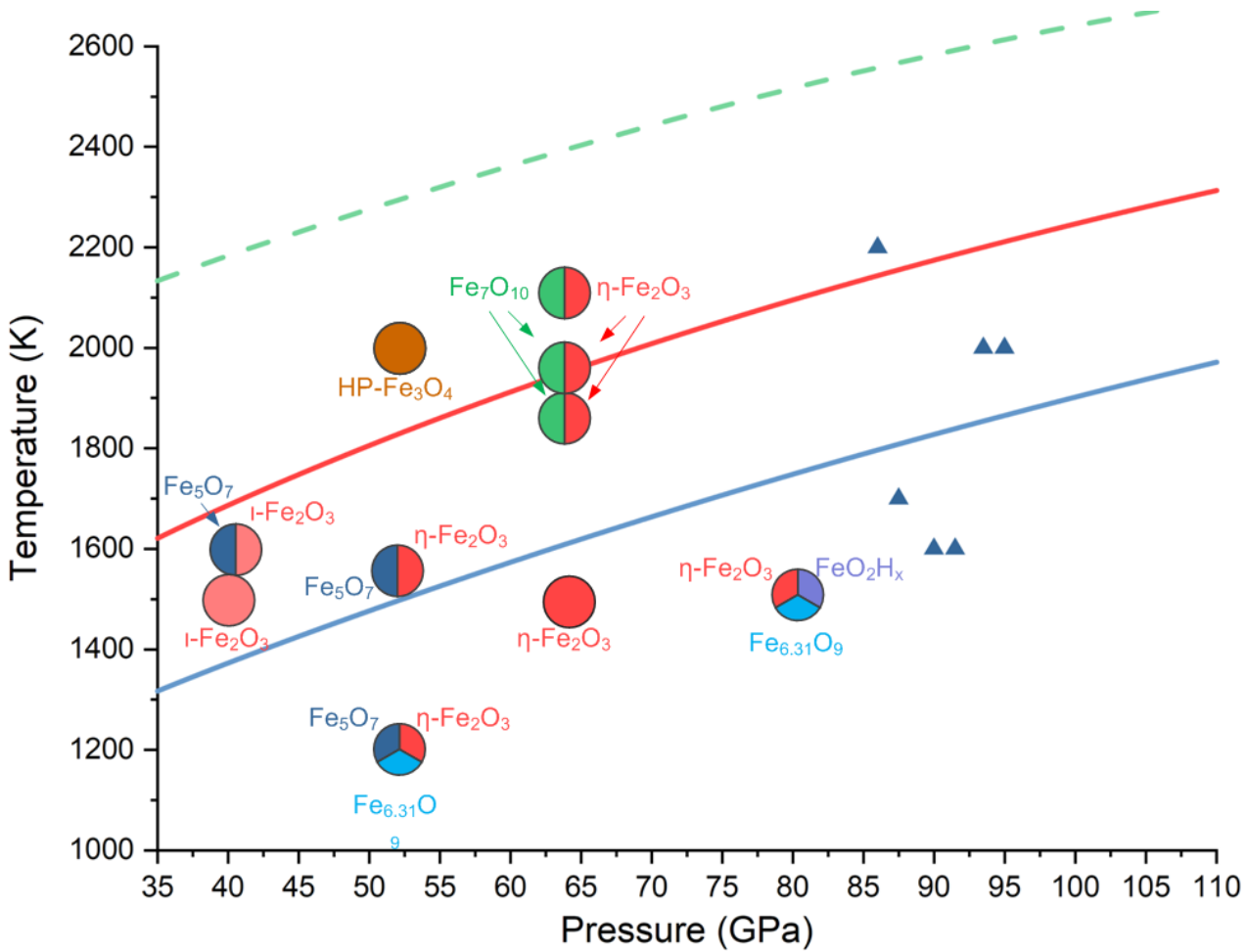

*Fig. 1. **Experimental conditions of HP-HT treatment of FeOOH and the results of phase analysis.** Dashed green line – Earth's mantle geotherm (37). Solid red and blue lines – calculated lowest temperatures across the slabs sinking at the 1 cm/year and 2 cm/year vertical speeds, correspondingly (for details see Supplementary Information). The position of filled circles represents P-T conditions of FeOOH treatment; different colors represent phases synthesized at these PT conditions (sectors size is arbitrary). Pressure error bars are within the size of the symbols. Blue solid triangles represent the synthesis conditions of $FeO_2H_x$ according to (38). We observed goethite's decomposition, which resulted in formation of previously known ι-$Fe_2O_3$, η-$Fe_2O_3$, HP-$Fe_3O_4$, and $Fe_5O_7$, and novel mixed-valence iron oxides $Fe_7O_{10}$ and $Fe_{6.32}O_9$. Formation of iron (II,III) oxides with release of oxygen upon treatment of $Fe_2O_3$ at pressures well above 50 GPa and heating over 2700 K was reported by Bykova et al. (18). We observed appearance of iron(II) bearing oxides (particularly, $Fe_5O_7$, $Fe_7O_{10}$, $Fe_{6.32}O_9$, etc.) at significantly lower pressures and temperatures than those required for decomposition of pure $Fe_2O_3$. This suggests that water is either directly involved or facilitates the reduction of ferric iron. Thus, water- and ferric-bearing phases become unstable upon heating at high pressure due to self-redox reactions.*

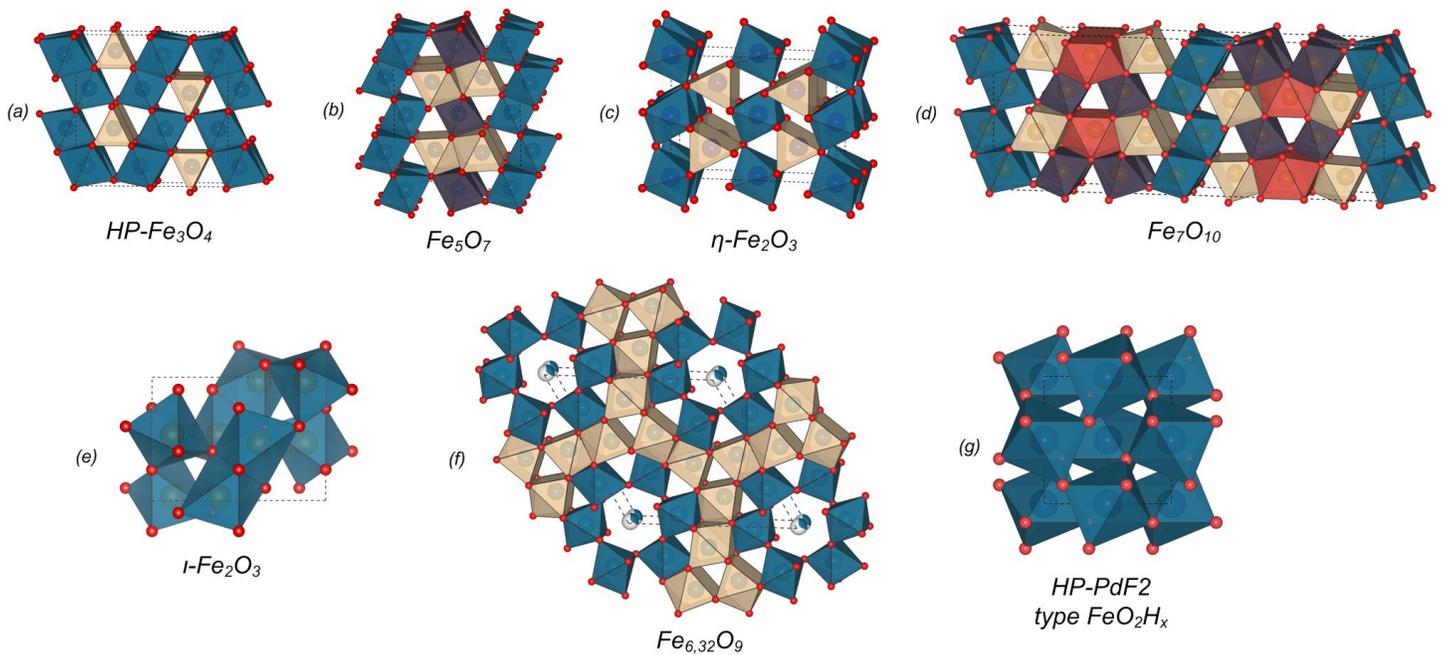

Fig. 2. Structures of iron oxides synthesized from goethite during HP-HT treatment as established by in situ single-crystal XRD. Blue spheres – iron atoms, red spheres – oxygen atoms. Building blocks are octahedra (blue and dark-purple) and trigonal prisms (pale yellow and orange). (a) HP-$Fe_3O_4$ (Bbmm) has the $CaTi_2O_4$ structural type, where sharing edges $FeO_6$ octahedra are ordered in a zigzag motif and interconnected via stacking layers of triangular prisms sharing their bases; (b) $Fe_5O_7$; (c) post-perovskite η-$Fe_2O_3$ ; (d) novel $Fe_7O_{10}$ (η-$Fe_2O_3$ and $Fe_7O_{10}$ are members of the homologous series $FeO \cdot mFe_2O_3$ (19), in there structures prisms are connected through common triangular faces, while octahedra are connected only via shared edges); (e) ι-$Fe_2O_3$ with a $Rh_2O_3$-II structure type(18, 19); (f) The structure of novel $Fe_{6.32}O_9$ can be described as hexagonal hollandite-like, with chains of apices-sharing $FeO_6$ octahedra forming hexagonal channels that are partially occupied by iron cations. A three-dimensional framework of mono-caped prisms separates the channels. (g) $FeO_2H_x$ (Pa-3) with the HP-$PdF_2$-type structure. The apices-sharing $FeO_6$ octahedra form a framework in which the shortest O-O distance is 2.267(5) Å at 82(1) GPa. For details of crystal structures of $Fe_7O_{10}$, $Fe_{6.32}O_9$, and $FeO_2H_x$ see Supplementary Materials.

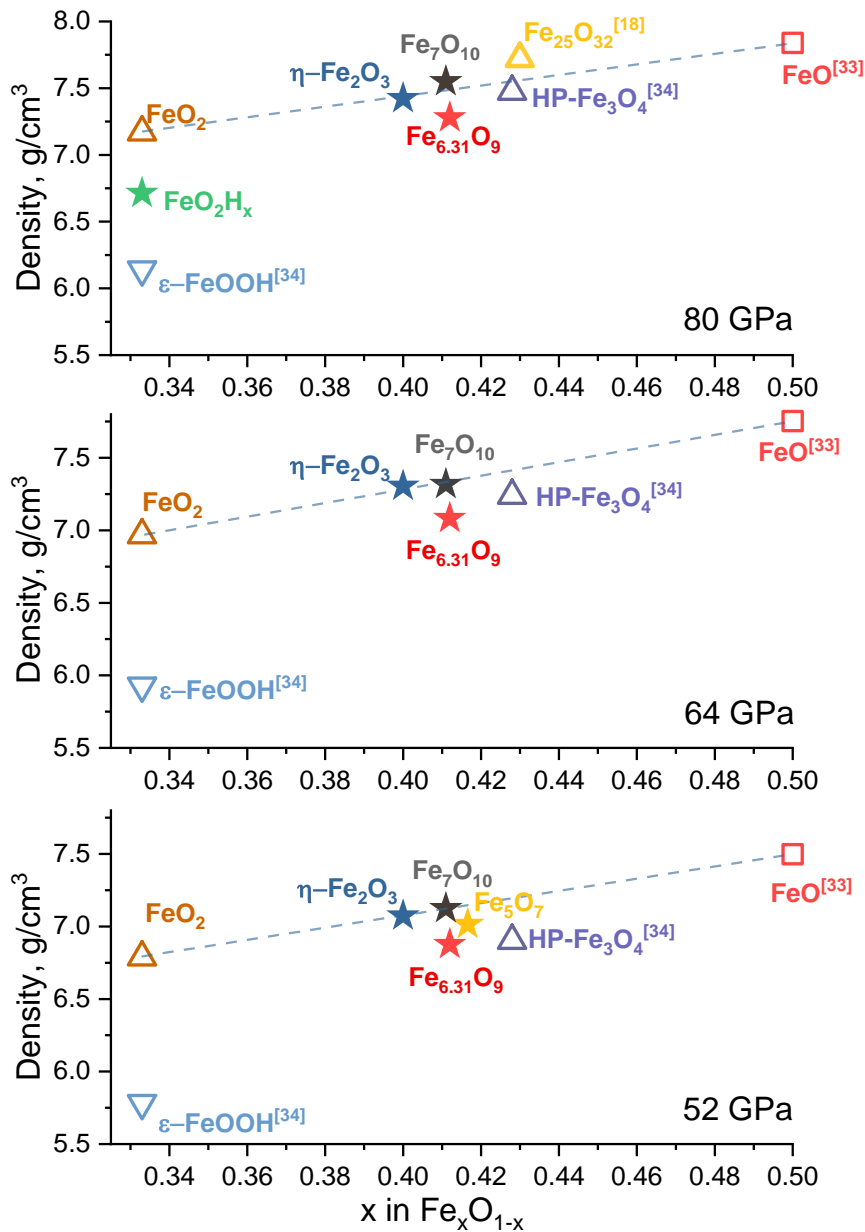

*Fig. 3. The **density of various iron-oxygen compounds as a function of their chemical composition at 52, 64, and 80 GPa and 300K**. Solid stars represent iron oxides obtained in present work, open symbols – the literature data. Error bars are less than the size of symbols. -Open purple triangles – densities of HP-$Fe_3O_4$ calculated from EOS (25). Open yellow triangle – the density of $Fe_{25}O_{32}$ at 80.1(5) GPa reported in (19). The density of η-$Fe_2O_3$ reported in (19) that found in this work (blue stars). Red open squares – the density of cubic FeO as calculated from the reported EOS (24). Open blue inverted triangles – the density of ε-FeOOH calculated from the EOS (22). Dashed blue lines are guides for eyes showing the trend of the density-stoichiometry dependence for iron oxides and connecting FeO and $FeO_2$ end-members. Densities of previously unknown iron oxides synthesized in present work do not deviate from this trend significantly that allows us to conclude that these iron oxides are nominally anhydrous. This highlights the dehydration/dehydrogenation of FeOOH during its decomposition.*

**References**

1. D. E. Canfield, M. T. Rosing, C. Bjerrum, Early anaerobic metabolisms. *Philos. Trans. R. Soc. B Biol. Sci.* **361**, 1819–1834 (2006).
2. S. A. Crowe *et al.*, Atmospheric oxygenation three billion years ago. *Nature*. **501**, 535–538 (2013).
3. O. Nebel *et al.*, Geological archive of the onset of plate tectonics. *Philos. Trans. R. Soc. A Math. Phys. Eng. Sci.* **376**, 20170405 (2018).
4. U. Söderlund *et al.*, Timing and tempo of the Great Oxidation Event. *Proc. Natl. Acad. Sci.* **114**, 1811–1816 (2017).
5. D. J. Frost, C. A. McCammon, The redox state of Earth's mantle. *Annu. Rev. Earth Planet. Sci.* **36**, 389–420 (2008).
6. M. M. Hirschmann, Water, melting, and the deep Earth H2O cycle. *Annu. Rev. Earth Planet. Sci.* **34**, 629–653 (2006).
7. D. R. Bell, G. R. Rossman, The role of Earth's mantle : *Science (80-. ).* **255**, 1391–1397 (1992).
8. K. Litasov, E. Ohtani, Phase relations and melt compositions in CMAS-pyrolite-$H_2O$ system up to 25 GPa. *Phys. Earth Planet. Inter.* **134**, 105–127 (2002).
9. S. D. Jacobsen, J. R. Smyth, *Earth's Deep Water Cycle* (2006), vol. 168.
10. H. K. Mao *et al.*, When water meets iron at Earth's core-mantle boundary. *Natl. Sci. Rev.* **4**, 870–878 (2017).
11. M. Nishi, Y. Kuwayama, J. Tsuchiya, T. Tsuchiya, The pyrite-Type high-pressure form of FeOOH. *Nature*. **547**, 205–208 (2017).
12. Q. Hu *et al.*, FeO2 and FeOOH under deep lower-mantle conditions and Earth's oxygen–hydrogen cycles (2016), doi:10.1038/nature18018.
13. Q. Hu *et al.*, Dehydrogenation of goethite in Earth's deep lower mantle. *Proc. Natl. Acad. Sci.* **114**, 201620644 (2017).
14. A. M. Dziewonski, D. L. Anderson, Preliminary reference Earth model. *Phys. Earth Planet.*

*Inter.* **25**, 297–356 (1981).

15. T. Katsura *et al.*, Adiabatic temperature profile in the mantle. *Phys. Earth Planet. Inter.* **183**, 212–218 (2010).

16. M. A. Eberle, O. Grasset, C. Sotin, A numerical study of the interaction between the mantle wedge, subducting slab, and overriding plate. *Phys. Earth Planet. Inter.* **134**, 191–202 (2002).

17. N. P. Butterworth *et al.*, Geological, tomographic, kinematic and geodynamic constraints on the dynamics of sinking slabs. *J. Geodyn.* **73**, 1–13 (2014).

18. E. Ito *et al.*, Determination of high-pressure phase equilibria of $Fe_2O_3$ using the Kawai-type apparatus equipped with sintered diamond anvils. *Am. Mineral.* **94**, 205–209 (2009).

19. E. Bykova *et al.*, Structural complexity of simple $Fe_2O_3$ at high pressures and temperatures. *Nat. Commun.* **7**, 10661 (2016).

20. M. L. Foo *et al.*, Synthesis and characterization of the pseudo-hexagonal hollandites $ALi_2Ru_6O_{12}$(A=Na, K). *J. Solid State Chem.* **179**, 941–948 (2006).

21. Q. Hu *et al.*, $FeO_2$ and FeOOH under deep lower-mantle conditions and Earth's oxygen–hydrogen cycles. *Nature*. **534**, 241–244 (2016).

22. A. E. Gleason, C. E. Quiroga, A. Suzuki, R. Pentcheva, W. L. Mao, Symmetrization driven spin transition in ε-FeOOH at high pressure. *Earth Planet. Sci. Lett.* **379**, 49–55 (2013).

23. W. Xu *et al.*, Pressure-induced hydrogen bond symmetrization in iron oxyhydroxide. *Phys. Rev. Lett.* **111**, 1–5 (2013).

24. R. A. Fischer *et al.*, Equation of state and phase diagram of FeO. *Earth Planet. Sci. Lett.* **304**, 496–502 (2011).

25. E. Greenberg *et al.*, High-pressure magnetic , electronic , and structural properties of M $Fe_2O_4$ ( M = Mg , Zn , Fe ) ferric spinels. **195150**, 1–13 (2017).

26. D. E. Canfield, *Oxygen: a four billion year history* (Princeton Uni Press, 2014), vol. 51.

27. D. P. Dobson, J. P. Brodholt, Subducted banded iron formations as a source of ultra-low velocity zones at the core-mantle boundary. *Nature*. **434**, 371–374 (2005).

28. D. V. Bekaert *et al.*, Archean kerogen as a new tracer of atmospheric evolution: Implications for dating the widespread nature of early life. *Sci. Adv.* **4**, 1–9 (2018).

29. G. Avice *et al.*, Evolution of atmospheric xenon and other noble gases inferred from Archean to Paleoproterozoic rocks. *Geochim. Cosmochim. Acta*. **232**, 82–100 (2018).

30. G. Avice, B. Marty, R. Burgess, The origin and degassing history of the Earth's atmosphere revealed by Archean xenon. *Nat. Commun.* **8**, 1–9 (2017).

31. I. Kantor *et al.*, BX90: A new diamond anvil cell design for X-ray diffraction and optical measurements. *Rev. Sci. Instrum.* **83** (2012), doi:10.1063/1.4768541.

32. J. A. Phys, Pressure calibration of diamond anvil Raman gauge to. **043516** (2006), doi:10.1063/1.2335683.

33. Y. Fei *et al.*, Toward an internally consistent pressure scale (2007).

34. O. D. Rigaku, CrysAlisPro Software System, Version 1.171. 38.41 l, Rigaku Coorporation (2015).

35. G. M. Sheldrick, Crystal structure refinement with SHELXL. *Acta Crystallogr. Sect. C Struct. Chem.* **71**, 3–8 (2015).

36. E. Bykova, thesis, Bayreuth (2015).

37. T. Katsura, A. Yoneda, D. Yamazaki, Adiabatic temperature profile in the mantle. **183**, 212–218 (2010).

38. J. Liu *et al.*, Hydrogen-bearing iron peroxide and the origin of ultralow-velocity zones. *Nature*. **551**, 494–497 (2017).